\documentclass[11pt,twoside]{article}
\usepackage{./asp2014}

\aspSuppressVolSlug
\resetcounters

\begin{document}

\title{The viscous decretion disk model of the classical Be star $\beta$ CMi revisited}
\author{R.~Klement,$^{1,2}$ A.C.~Carciofi,$^3$ and T.~Rivinius$^1$}
\affil{$^1$European Southern Observatory, Alonso de Cordova 3107, Vitacura, Casilla 19001, Santiago, Chile; \email{robertklement@gmail.com}}
\affil{$^2$Astronomical Institute of Charles University, Charles University, V~Hole\v sovi\v ck\'ach 2, 180 00  Prague 8}
\affil{$^3$Instituto de Astronomia, Geof\'isica e Ci\^encias Atmosf\'ericas, Universidade de S\~ao Paulo, Rua do Mat\~ao 1226, Cidade Universit\'aria, 05508-090, S\~ao Paulo, SP, Brazil}

\paperauthor{Robert Klement}{robertklement@gmail.com}{}{European Southern Observatory}{}{Santiago}{}{}{Chile}
\paperauthor{Alex Cavalieri Carciofi}{}{}{Instituto de Astronomia, Geof\'isica e Ci\^encias Atmosf\'ericas}{Universidade de S\~ao Paulo}{S\~ao Paulo}{}{}{Brazil}
\paperauthor{Thomas Rivinius}{}{}{European Southern Observatory}{}{Santiago}{}{}{Chile}

\begin{abstract}
We revisit the viscous decretion disk (VDD) model of the classical Be star $\beta$ CMi as presented by \citet{klement} using an updated version of the radiative transfer code {\ttfamily HDUST}. A software bug was causing the mean intensities to be slightly underestimated in the equatorial region of the disk, with small but detectable effects on the disk temperature and emergent spectrum. The new model fixes an unexplained feature of the original model, which was able to reproduce the observations only when considering a dual density behavior: a steep density fall-off in the very inner parts of the disk followed by a shallower density profile. The new model is able to reproduce all the observables reasonably well using a single power-law for the density profile throughout the whole disk, as predicted by the VDD model. All the other original conclusions, most importantly the reported truncation of the disk at the distance of 35 stellar equatorial radii from the central star, remain unchanged.
\end{abstract}

\section{Introduction}\label{intro}
The viscous decretion disk (VDD) model of $\beta$~CMi presented by \citet{klement} remains one of the most detailed physical models of an individual Be star to this date. In that study, VDD model predictions implemented in the radiative transfer code {\ttfamily HDUST} \citep{hdust1} were confronted with a large multi-technique dataset including spectral energy distribution (SED) measurements covering the interval from the ultraviolet (UV) to radio wavelengths, spectral line profiles, linear polarimetry and several optical/near-IR interferometric datasets. The modeling revealed a very good agreement with the observations when using a simple VDD model with a power law density fall-off. Complications arose for the observables originating in the very inner parts of the disk ($\lesssim 5$\,$R_\text{e}$, where $R_\text{e}$ is the stellar equatorial radius), which needed a lower density model to be reproduced. As for the outermost parts of the disk, observable at radio wavelengths, the modeling revealed signs of the disk being truncated at the distance of $35^{+10}_{-5}$\,$R_\text{e}$ from the central star. The disk truncation was interpreted to arise due to a previously undetected binary companion tidally influencing the outer parts of the Be disk. Subsequent radial velocity analysis of the H$\alpha$ line brought the detection of the orbital period of $170.36 \pm 4.27$ days (Dulaney et al., in prep.) and therefore a confirmation of the presence of the companion. The orbital period of a particle at the truncation radius is very close to the 3:2 resonance with the derived orbit.

Since the publication of the study of \citet{klement}, a software bug was found in the {\ttfamily HDUST} code version 2.02, causing the resulting mean intensities to be slightly underestimated in the equatorial region, with small but detectable effects in the disk temperature and emergent spectrum. The new, corrected version ({\ttfamily HDUST} 2.10) causes the temperature to be slightly lower and the level populations more neutral. The changes in the resulting synthetic observables are mostly minor: a decrease in polarization percentage, line emission, and the IR flux levels. However, the differences between the old and new versions get larger with the increasing rotational velocity of the central star, because the emerging flux becomes strongly dependent on stellar latitude. $\beta$~CMi was found to rotate very close to (or even at) the critical rate, therefore the results that were published previously need to be revisited.

\section{Observations}
We use the same set of observations as in the original paper. For details see Sect. 2 of \citet{klement}.

A minor change regards the observed flux density from APEX\footnote{This publication is based on data acquired with the Atacama Pathfinder Experiment (APEX). APEX is a collaboration between the Max-Planck-Institut fur Radioastronomie, the European Southern Observatory, and the Onsala Space Observatory}/LABOCA. New reduction using the updated Crush\footnote{\url{http://www.submm.caltech.edu/~sharc/crush/index.html}} data reduction software version 2.32-1 results in the observed flux density of $38.6 \pm 3.1$\,mJy. The value of the originally published flux density was $33.5 \pm 6.5$\,mJy, so the result remains consistent also within the smaller errors of the new reduction. We also use a more conservative estimate of the flux error from the CARMA array flux measurement at 3\,mm: we use a 10\% relative error of 0.96\,mJy as opposed to previously assumed 0.60\,mJy, which was most probably underestimated.

\section{Results}
Here we discuss the changes to the model when using the corrected version of {\ttfamily HDUST}. We use the parametric VDD model, as described in Sect.~3 of \citet{klement}. In that model, the radial density profile of the disk is a simple power law in the form $\rho(r) = \rho_0 r^{-n}$, where $r$ is the distance from the star, $\rho_0$ is the disk base density and $n$ is the power law density exponent. This model has three free parameters: $\rho_0$, $n$ and the physical extent of the disk (outer disk radius $R_\text{out}$). We recall that $n=3.5$ is the canonical value for isothermal Be disks in steady state, that are not affected by outside influences, such as a close orbiting companion. Note that second-order effects, currently not taken into account, may cause a deviation from $n=3.5$ even in a steady-state, isolated disk. One such effect is cooling by heavier elements. When the disks are in steady state, the values of $n$ actually seem to range from 3.0 to 3.5 \citep{2016MNRAS.tmp.1532V}. By letting the value of $n$ (the same throughout the whole disk) to vary in our model, we allow for the possibility that the disk may be growing ($n>3.5$) or dissipating \citep[$n<3.0$, see][]{haubois1}. Also, we account for possible accumulation and truncation effects due to an orbiting secondary companion. These effects cause $n$ to get lower than 3.5 inwards of the truncation radius and much higher than 3.5 outwards of the truncation radius. The impact of the accumulation effect gets larger with decreasing orbital period, decreasing viscosity, and increasing binary mass ratio \citep{panoglou}.

\begin{figure}[!t]
\begin{centering}
\includegraphics[width=1.\textwidth]{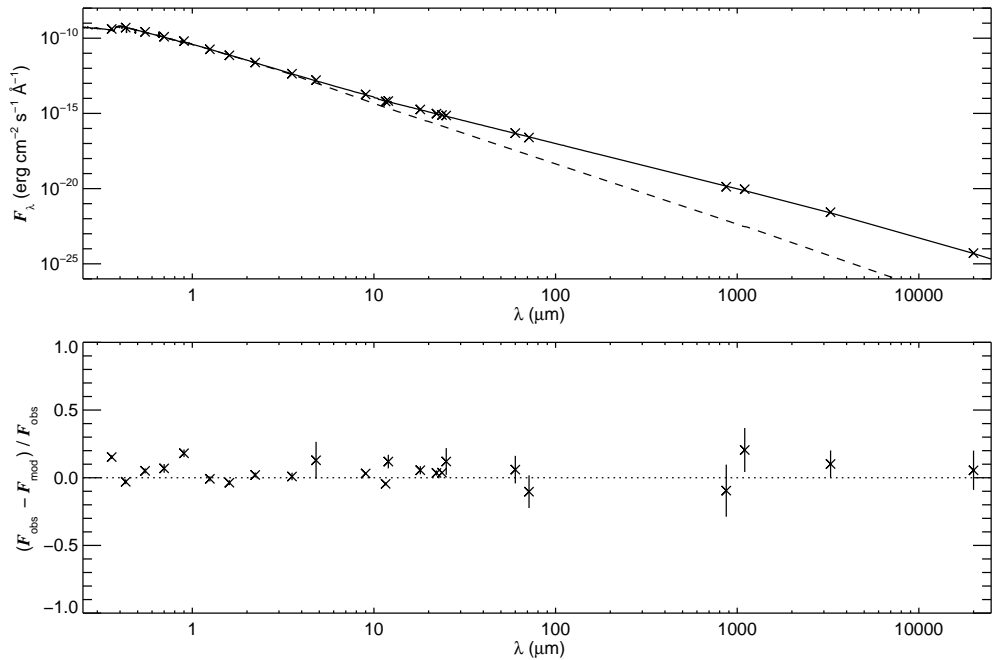}
\caption{\textit{Upper: } The best-fit parametric model with $\rho_0 = 2.0 \times 10^{-12}$ g\,cm$^{-3}$, $n = 2.9$ and $R_\text{out} = 35$\,$R_\text{e}$ (solid line) reproducing the visual, IR and radio SED (crosses). The purely photospheric SED is plotted as a dashed line. \textit{Lower: } Resi\-duals of the best-fit model.}
\label{sed}
\end{centering}
\end{figure}

Below we follow the same order of comparing the synthetic observables to the observations as in the original paper. We discuss the changes to the SED structure, the synthetic line profiles, linear polarimetry and finally the optical/near-IR interferometry.

\subsection{SED}
To perfectly reproduce the full SED, combination of two models had to be originally used. While the inner parts followed a steep density fall-off with $n=3.5$, the remainder of the disk followed a shallower profile with $n=3.0$. Furthermore, the observed flux density at 2\,cm could be reproduced only when using a disk truncated at 35\,$R_\text{e}$ from the central star.

The new model reveals that the dual density behavior was an artifact caused by the underestimation of the model fluxes in the near-IR. Using the corrected version, the full SED is now very well reproduced by a single power-law with $n=2.9$ (Fig.~\ref{sed}). The conclusion regarding the disk truncation at 35\,$R_\text{e}$ from the central star and the interpretation as being caused by an unseen secondary companion remains unchanged.

\articlefigure[width=.72\textwidth]{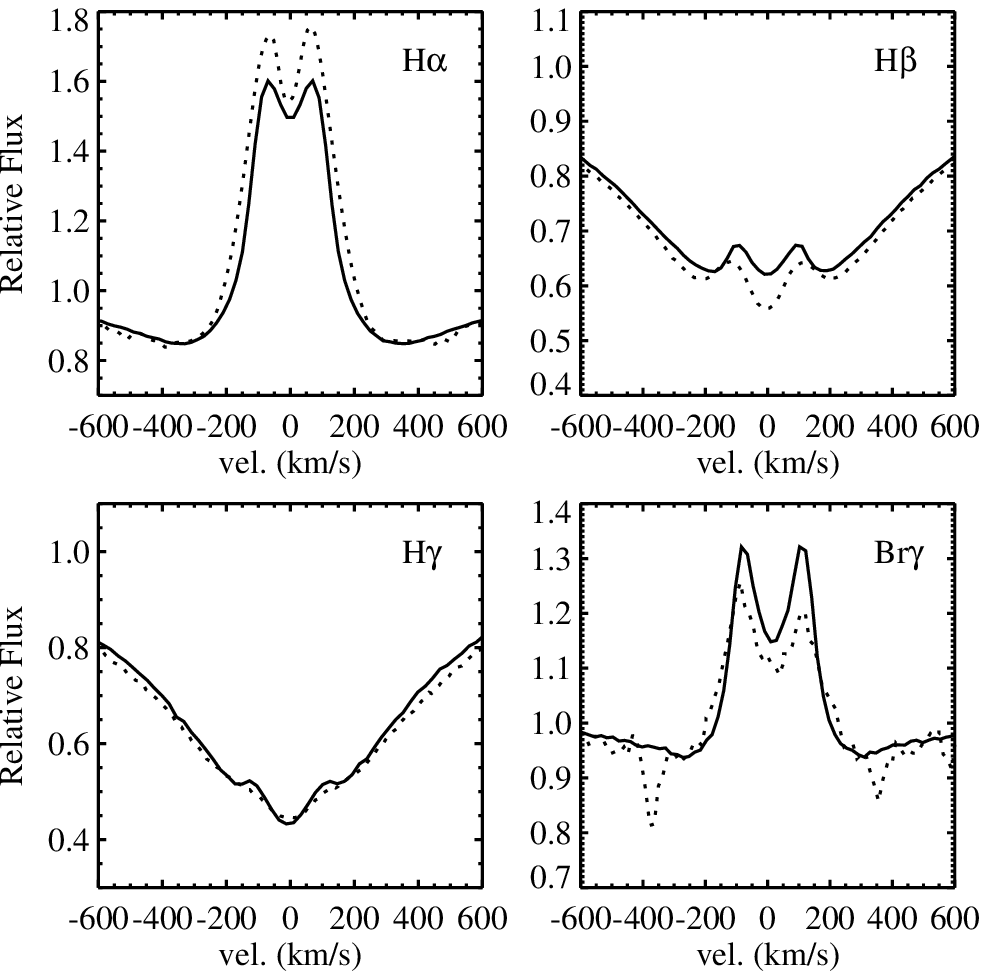}{lines}{Hydrogen line profiles for the parametric model with $n=2.9$ (solid lines) compared to the averaged observed line profiles (dotted lines).}

\subsection{Hydrogen emission line profiles}
Similarly to the SED, in the original study the emission line profiles could not be reproduced by a model with a single value of $n$ throughout the whole disk. The lines H$\beta$ and H$\gamma$, originating closer to the central star needed the $n=3.5$  model, while the lines H$\alpha$ and Br$\gamma$, originating over much larger volume of the disk, were well reproduced with $n=3.0$. This result further corroborated the fact that there seemed to be a dual density behavior throughout the disk.

The new model reveals that that is no longer the case. All lines are now reproduced reasonably well by the $n=2.9$ model, which best fits the observed SED. There are some small discrepancies, such as the amount of emission in the H$\alpha$ line being slightly underestimated, while being overestimated for the H$\beta$ and Br$\gamma$ lines. However, the new result is still a major improvement over the original fit to the lines when using a single value of $n$.

\articlefigure[width=.72\textwidth]{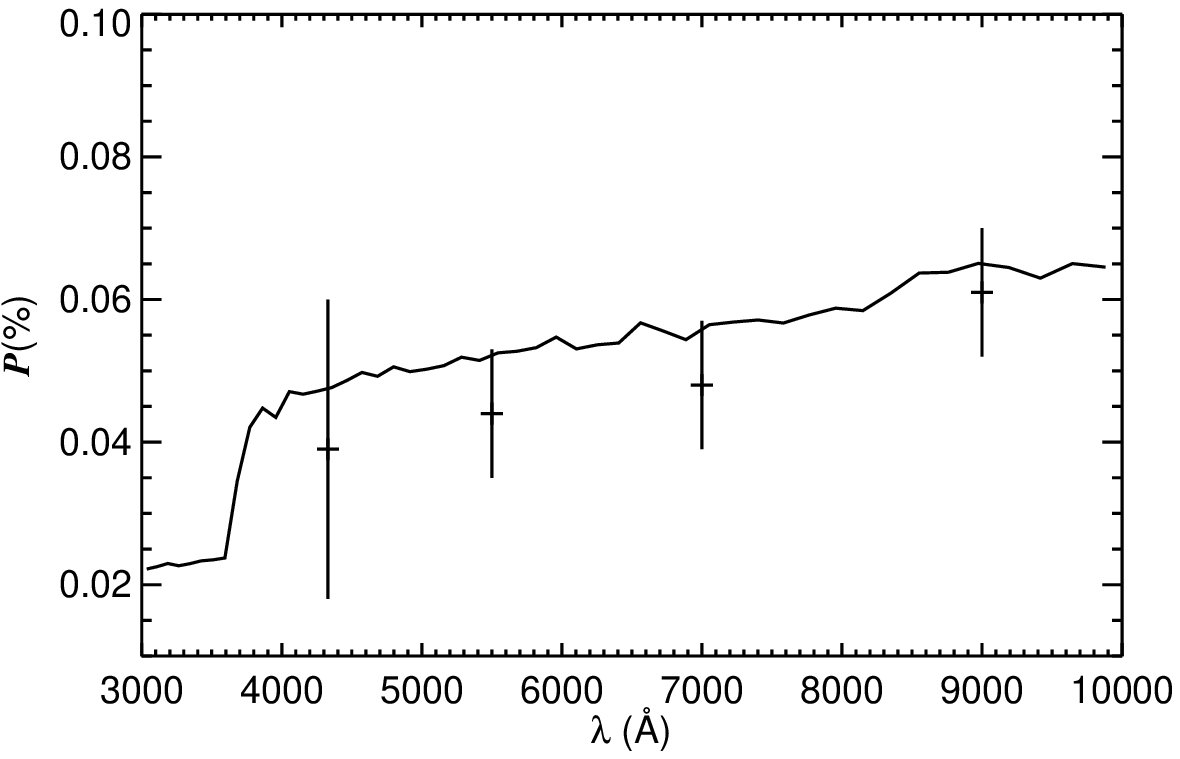}{pol}{Spectral shape of polarization of the parametric model with $n=2.9$ (solid line). The OPD measurements are plotted as plus signs with error bars.}

\subsection{Optical polarization}
Analysis of the observed linear polarization and the comparison with our models again showed a result consistent with what was stated above for the SED and emission line profiles. Since the optical polarization originated in the very inner parts of the disk, a steeper density profile ($n=3.5$) was needed in order to match the observed polarization level.

The situation changed after recomputing the model with the fixed version of the code. Even the $n=2.9$ model now reproduces the observed polarization level within the observed error bars, while the slope of the polarization in the Paschen continuum agrees perfectly with the observations. The result that the slope of the optical polarization of tenuous disks becomes negative with increasing rotation rate \citep[Fig.~8 of][]{klement} and that $\beta$~CMi rotates very close to critical ($W\gtrsim0.98$) remains unchanged.

\section{Optical/near-IR interferometry}
The changes to the interferometric part of the original study are mostly minor. In Table 7 of the original paper, we compared the resulting $\chi^2$ values of the 2 parametric models with $n=3.5$ and $n=3.0$ respectively. The recomputed values can be found in Table~1, in which we also include the $n=2.9$ model, which best reproduces the observed SED. The fit of the $n=2.9$ model to the AMBER data unfortunately shows a slight decline in agreement as compared to the $n=3.0$ model. The resulting $\chi^2$ values of the fits to the rest of the data have remained very similar. This is not surprising since most of these observables originate over large areas of the disk.

\section{Conclusions}
We have presented the changes to the VDD model of $\beta$~CMi, that occured after fixing an issue with the {\ttfamily HDUST} code. Unlike the original conclusion of \citet{klement}, the whole multi-technique and multi-wavelength dataset is now well reproduced by a parametric model with a single power-law density exponent ($n=2.9$) throughout the whole disk. This improves the agreement between the VDD model predictions and the multi-technique observations as compared to the results presented in the original study. Therefore the VDD model is further established as the correct one for the disks of Be stars. 

The other conclusions of the original study remain unchanged, most importantly, the radio flux measurement at 2\,cm still shows clear truncation effects with the best-fit disk size being 35 stellar equatorial radii. The subsequent detection of an orbital period of $170.36 \pm 4.27$ days from radial velocity analysis of the H$\alpha$ line confirms that it is indeed a previously unknown binary companion that is the cause of the disk truncation.

\begin{table}[!t]
\caption{Model fits to the interferometric measurements.}
\smallskip
\begin{center}
{\small
\begin{tabular}{lcccc}  
\tableline
\noalign{\smallskip}
Instrument & Spectral band & $\chi^2$ ($n=2.9$) & $\chi^2$ ($n=3.0$) & $\chi^2$ ($n=3.5$) \\
\noalign{\smallskip}
\tableline
\noalign{\smallskip}
NPOI & $R$ (H$\alpha$) & 1.37 & 1.37 & 1.92 \\
MIRC & $H$ & 2.17 & 2.24 & 2.39 \\
FSU-A & $K$ & 0.89 & 0.87 & 0.83 \\
AMBER & $K$ (Br$\gamma$) & 1.52 & 1.20 & 4.30 \\
CLIMB & $K$' & 1.48 & 1.53 & 1.64 \\
MIDI & $N$ & 1.83 & 1.86 & 1.95 \\
\noalign{\smallskip}
\tableline\
\end{tabular}
}
\end{center}
\end{table}

\acknowledgements The research of R.K. was supported by grant project number 1808214 of the Charles University Grant Agency (GA UK). A.C.C. acknowledges support from CNPq (grant 307594/2015-7) and FAPESP (grant 2015/17967-7).

\bibliography{biblio}  


\end{document}